\documentclass[showpacs]{article}
\usepackage{graphicx}

\begin{document}

\begin{center}
{\Large \textbf{Further evidence for linearly-dispersive Cooper
pairs}}
\end{center}

\begin{center}
M. de Llano$^{a,b}$ and J.J. Valencia$^{b,c}$*\\

$^{a}$Texas Center for Superconductivity, University of Houston,
Houston, TX
77204, USA\\
$^{b}$Instituto de Investigaciones en Materiales, UNAM, 04510
M\'{e}xico,
DF, Mexico\\
$^{c}$Universidad Aut\'{o}noma de la Ciudad de M\'{e}xico,
M\'{e}xico, San Lorenzo Tezonco, 09790 M\'{e}xico, DF, Mexico
\end{center}

\begin{center}
\textbf{Abstract}
\end{center}
A recent Bose-Einstein condensation (BEC) model of several cuprate
superconductors is based on bosonic Cooper pairs (CPs) moving in
3D with a \textit{quadratic} energy-momentum (dispersion)
relation. The 3D BEC condensate-fraction \textit{vs}. temperature
formula poorly fits penetration-depth data for two cuprates in the
range $1/2 < T/T_{c}\leq 1$ where $T_{c}$ is the BEC transition
temperature. We show how these fits are dramatically improved
assuming cuprates to be quasi-2D, and how equally good fits obtain
for conventional 3D and quasi-1D nanotube superconducting data,
provided the correct \textit{linear }CP dispersion is assumed in
BEC at their assumed corresponding dimensionalities. This is
offered as additional concrete empirical evidence for
linearly-dispersive pairs in another recent BEC scenario of
superconductors within which a BCS condensate turns out to be a
very special case.


\begin{center}
\large{\textbf{Introduction}}
\end{center}

A Bose-Einstein condensation (BEC) model was applied by Rosencwaig
in Ref. \cite{Rosencwaig} to address seven cuprate superconductors
(SCs) with transition temperatures $T_{c}$ at optimal doping
ranging from 22 K to 133 K. These are: La$_{2-x}$Sr$_{x}$CuO$_{4}$
(LSCO), Nd$_{2-x}$Ce$_x$CuO$_4$ (NCCO), YBa$_{2}$Cu$_{3}$O$_{7-y}$
Y123, Bi$_{2}$Sr$_{2}$CaCu$_2$O$_{8-y}$ Bi2212,
Bi$_2$Sr$_2$Ca$_{2}$Cu$_{3}$O$_{10-y}$ (Bi2223),
HgBa$_{2}$CaCu$_{2}$O$_{7-y}$ (Hg1212) and
HgBa$_{2}$Ca$_{2}$Cu$_{3}$O$_{9-y}$ (Hg1223). His starting point
is the well-known fact that BEC in an ideal Bose gas occurs below
temperatures $T$ such that the thermal wavelength $\lambda \equiv
h/\sqrt{2\pi m_{B}k_{B}T}$ becomes larger than the average
interbosonic separation, with $m_B$ the boson mass, and $h$,
$k_{B}$ the Planck and Boltzmann constants, respectively. More
exactly, BEC sets in whenever
\begin{equation}
n_{B}\lambda ^{3}> 2.612  \label{1}
\end{equation}
where $n_{B}$ is the boson number density, and $\lambda$ is taken
as the bosonic quasiparticle diameter. This leads to a critical
temperature $T_{c}$ given by the familiar formula
\begin{equation}
T_{c}={\frac{2\pi \hbar ^{2}n_{B}^{2/3}}{(2.612)^{2/3}m_{B}k_{B}}}\simeq {%
\frac{3.31\hbar ^{2}n_{B}^{2/3}}{m_{B}k_{B}}} \propto n_{B}^{2/3}
\label{2}
\end{equation}
of conventional BEC theory. He identifies an interaction distance
with$\lambda $, which thus becomes $T$-independent, while the
number-density $n_{B}$ of weakly-interacting ``preformed''
electron or hole pairs acquires a $T$-dependence. Associated with
(\ref{2}) is the BE condensate fraction
\begin{equation}
N_{0}(T)/N_{0}(0)=\left\{
\begin{array}{c}
1-(T/T_{c})^{3/2}\hbox{ \ \ for }T\leq T_{c} \\
0\hbox{\ \ \ otherwise}
\end{array}
\right.   \label{3}
\end{equation}%
where $N_{0}(T)$ is the number of bosonic pairs at temperature
$T$\ in the
lowest-energy state with (total, or) center-of-mass momentum wavenumber $%
K=0, $ and $N_{0}(0)$\ is that same number at $T=0$.

\begin{center}
\large{\textbf{Cooper pair dispersion}}
\end{center}

All of this assumes three dimensions (3D) and that boson
excitation energies are given by
\begin{equation}
\varepsilon _{K}=\hbar ^{2}K^{2}/2m_{B}  \label{4}
\end{equation}%
with as would hold if the composite bosons moved \textit{in vacuo}
such as, say, a deuteron of mass $m_{B}=m_{p}+m_{n}$\ in empty
space with $m_{p}$ and $m_{n}$ being the proton and neutron
masses. However, in the presence of the Fermi sea of the other
single\ charge carriers the bosonic ``dispersion relation''
becomes \textit{linear }\cite{Schrieffer64}\cite{fu}\ in leading
order rather than quadratic as in (\ref{4}). Ref.
\cite{Schrieffer64} first mentioned, and Refs.
\cite{FW}\cite{PhysicaC} later discussed this linearity in detail.
It is associated with the original Cooper pair (CP)\ problem
\cite{Cooper}\ of two electrons (or holes) above (or below) the
Fermi surface of the remaining system electrons. It was also found
in a more general view of CPs in Refs. \cite{Honolulu}\cite{EPJD}
within a many-body Green's function formalism treating both
electron- and hole-pairs on an equal footing. For either ordinary
or generalized CPs, the leading term in the $K$-expansion is
linear. This linearly-dispersive ``moving CP'' object is often
confused in the literature\ with the more common
Anderson-Bogoliubov-Higgs (ABH) \cite{ABH} (Ref. \cite{bts} p. 44) \cite%
{Higgs} collective excitation which is also linear in leading
order, but which is just the sound mode of the many-fermion
system.\ By contrast, in a many-boson system these two modes, the
``particle'' and ``sound'' modes, are apparently identical
\cite{Gavoret}\cite{Hohenberg}.

A particularly clear example comparing linear and quadratic
dispersion is perhaps the analytical result of Ref. \cite{PRB2000}
in 2D for an attractive delta potential assumed between electrons.
This interfermion interaction mimics the net effect of Coulomb
repulsion plus attractive, say, electron-phonon interactions. The
2D delta potential well, which otherwise
supports an infinite number of bound levels, is imagined ``regularized'' %
\cite{GT} to support a single bound level of energy $-B_{2}$ as occurs \cite%
{IJTP}, e.g., with the two-parameter Cooper/BCS \cite{Cooper}\cite{BCS}%
\ model interelectronic interaction. Miyake used this interaction to obtain %
\cite{Miyake}\ both the zero-temperature BCS gap $\Delta $ and the
chemical potential $\mu $\ analytically in terms of $B_{2}.$ Since
the regularized delta well turns out to be infinitesimally weak,
its $0^{+}$ strength can be eliminated \cite{PRB2000}\ in favor of
$B_{2}$ which then plays the role of coupling constant, with
$0\leq B_{2}<\infty $ spanning weak to strong coupling. Instead of
(\ref{4}), a more general analytical expression found in Ref.
\cite{PRB2000} that includes the Fermi sea is
\begin{equation}
\varepsilon _{K} =\frac{2}{\pi }\hbar v_{F}K+\left[ 1-\left\{
2-\left( \frac{4}{\pi }\right) ^{2}\right\}
\frac{E_{F}}{%
B_{2}}\right] \frac{\hbar ^{2}K^{2}}{2(2m_{e})}+O(K^{3})
\label{5}
\end{equation}%
where $v_{F}$ is the Fermi velocity defined through the Fermi
surface energy $E_{F}\equiv m_{e}v_{F}^{2}/2$ and $m_{e}$ is the
effective electron mass.
The leading term is linear, and \textit{only} in the vacuum limit ($%
v_{F}\rightarrow 0$, implying $E_{F}\rightarrow 0$) does it
precisely become the quadratic (\ref{4}) with $m_{B}=2m_{e}$
expected physically \textit{for any fixed coupling } $B_{2}$. Fig.
1 of Ref. \cite{PRB2000} exhibits the smooth crossover in 2D from
a purely linear to a purely quadratic form, as one increases
coupling and/or as one ``switches off'' the Fermi sea medium
(nonzero $E_{F}$)\ in which the pair propagates down to the pure
vacuum (zero $E_{F}$) medium. A very similar behavior was also
observed in 3D\ \cite{PhysicaC}, but only numerically.

\begin{center}
\large{\textbf{Bose-Einstein condensation}}
\end{center}

Expressions more general than (\ref{2}) and (\ref{3})\ but
reducing to them, are known \cite{PLA98} in any dimensionality
$d>0$ (integer or not) and for any dispersion relation
\begin{equation}
\varepsilon _{K}=C_{s}\,K^{s},\qquad \hbox{with}\quad s>0.
\label{6}
\end{equation}%
They are
\begin{equation}
T_{c}=\frac{C_{s}}{k_{B}}\left[ \frac{s\Gamma (d/2)(2\pi
)^{d}n_{B}}{2\pi ^{d/2}\Gamma (d/s)g_{d/s}(1)}\right] ^{s/d}\quad
 \propto \quad n_{B}^{s/d}  \label{7}
\end{equation}%
and%
\begin{equation}
N_{0}(T)/N_{0}(0)=\left\{
\begin{array}{c}
1-(T/T_{c})^{d/s}\hbox{ \ \ \ for }T\leq T_{c} \\
0\hbox{\ \ \ \ otherwise}.%
\end{array}%
\right.  \label{8}
\end{equation}%
Here\ $g_{\sigma }(1)$ are the Bose integrals which for $\sigma
>1$ coincide with the Riemann Zeta-function $\zeta (\sigma )$ and
diverge for $\sigma \leq 1$, $\Gamma (\sigma )$ is the gamma
function, and $n_{B}\equiv N/L^{d}$ is the $d$-dimensional boson
number density. The divergence of $g_{\sigma }(1)$ for $\sigma
\leq 1$ ensures from (\ref{7}) that $T_{c}\equiv 0$ for all $d\leq
s$, but that otherwise $T_{c}$ is nonvanishing. In $d=3$ and
quadratic dispersion $s=2$ and, if $C_{2}=\hbar ^{2}/2m_{B}$, (\ref{7}) and (%
\ref{8}) respectively become (\ref{2}) and (\ref{3}), as
$g_{3/2}(1)\equiv
\zeta (3/2)\simeq 2.612$. However, for $s=1$ BEC \textit{can }occur for all $%
d>1$. This coincides, fortuitously, with \textit{all} dimensions
where actual superconductors have been found to exist, down to the
quasi-one-dimensional organics
\cite{organometallics}\cite{jerome2}\ consisting of parallel
chains of molecules. As regards dimensionality,
therefore, the BEC picture contrasts sharply with the BCS scheme where $%
T_{c} $ is nonvanishing for all $d>0$ even though no exactly 1D
superconductors have been found to date. In fact, beautiful experiments \cite%
{Tinkham00}\cite{Tinkham01} with nanowires of different
thicknesses sputter-coated with an amorphous superconductor
($T_{c}\simeq 5.5K$) have shown how superconductivity is
extinguished for the smallest nanowire diameters interpreted as
approaching precisely 1D.

Although the creation/annihilation operators of BCS pairs do
\textit{not} obey the usual Bose commutation rules [see Eqs.
(2.11) to (2.13) of Ref. \cite{BCS}; see also p. 38 of Ref.
\cite{Schrieffer64}], CPs in fact
\textit{satisfy BE statistics.} Indeed, BCS pairs and CPs are \textit{%
distinct. }A BCS pair is defined with fixed total (or
center-of-mass)
momentum wavevector $\mathbf{K}\equiv \mathbf{k}_{1}+\mathbf{k}_{2}$ \textit{%
and }fixed relative-momentum wavevector $\mathbf{k}\equiv (\mathbf{k}_{1}-%
\mathbf{k}_{2})/2,\mathbf{\ }$whereas a CP is defined with fixed
$\mathbf{K}$ only, since a sum over $\mathbf{k}$ is implied in any
conceivable formulation\ of CPs\textit{. }This is because in the
thermodynamic limit an indefinitely large number of BCS pairs,
each with fixed momenta $\hbar \bf{k}_{1}$ and $\hbar \bf{k}_{2},$
correspond to different
relative momenta $\hbar \bf{k}$ but whose $\mathbf{k}_{1}$ and $%
\mathbf{k}_{2}$\ add vectorially to the \textit{same} total
$\mathbf{K}.$ These remarks apply even when only $\mathbf{K=0}$
pairs were considered in Ref. \cite{BCS}

\begin{center}
\large{\textbf{Results}}
\end{center}

Empirical evidence for the \textit{linearly-dispersive} nature of
CPs in BSCCO has been argued by Wilson \cite{Wilson03} as being
suggested by the scanning tunneling microscope conductance
scattering data obtained by Davis and coworkers
\cite{Hoffman02}\cite{McElroy03} in this cuprate. In Figure 1 we
present additional evidence, based on experimental data from
penetration-depth measurements\ in two 3D SCs \cite{Guimpel}\cite%
{Schawlow} and two quasi-2D cuprates \cite{Jacobs}\cite{Sonier},
as well as from gap measurements in a quasi-1D nanotube SC
\cite{Tang}. When plotted as a presumably universal ``normalized
order parameter'' the data\ depart substantially (at least in 3D
and 2D)\ from the BCS normalized gap order parameter, but are seen
to agree quite well, at least for $T> 0.5T_{c}$, with the
\textit{pure-phase} (only 2e- \textit{or} 2h-CP) BEC
condensate-fraction formula (\ref{8})$\ $for $d=3$, $2$ and $1$, \textit{%
provided one assumes }$s=1.$ For lower $T$'s, one can argue
\cite{PA03} that a \textit{mixed }BEC phase containing both 2e-
and 2h-CPs becomes more stable (i.e., has lower Helmholtz free
energy) so that the simple pure-phase formula (\ref{8})\ is no
longer strictly valid. Indeed, (\ref{8}) applies at all to the CPs
because in the binary boson-fermion gas mixture---for, say, a
Cooper/BCS model interaction forming the bosonic CPs with a maximum allowed %
\cite{Migdal}\cite{Blatt64} coupling of $\lambda =1/2$---only a
minuscule fraction ($<0.1\%$) \cite{SSC02}\ of the individual
fermion charge carriers are paired up into CPs, ensuring that a
substantial Fermi sea is still present. Such tiny fractions are
consistent with some very recent far-infrared charge-dynamics
measurements \cite{Hor}\cite{Hor02}\ in LSCO.

\begin{center}
\large{\textbf{Conclusions}}
\end{center}

To conclude, we have presented normalized order-parameter data
based on
penetration-depth and gap measurements that strongly suggest a \textit{%
linear }energy vs. center-of-mass-momentum (dispersion) relation
for Cooper pairs in various materials that can be viewed as 3D,
quasi-2D and quasi-1D superconductors (SCs). The linearity is a
manifestation of the Fermi sea background in which the pairs
propagate, as opposed to the quadratic relation of composite
bosons moving \textit{in vacuo. }It ensures that a BEC picture of
SCs is applicable over all dimensionalities in which SCs occur.

%FIGURA 1
\begin{figure}[t]
\begin{center}
\includegraphics[height=4.0in,width=4.0in,angle=90]{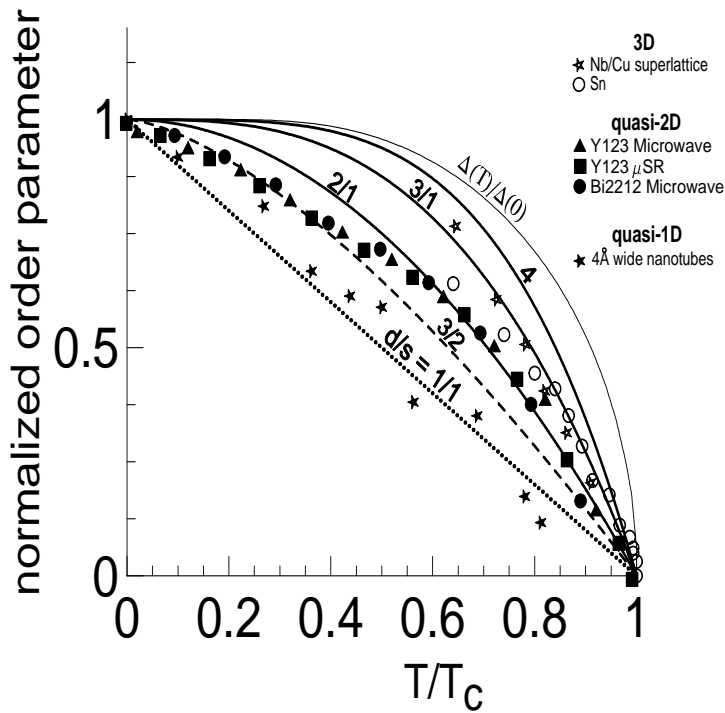}
\end{center}
\caption{BE condensate-fraction curves for bosons in $d=3$, $2$,
or $1$.} \label{}
\end{figure}

In Figure 1 we show BE condensate-fraction curves (in thick) $%
1-(T/T_{c})^{d/s}$ for bosons in $d=3$, $2$, or $1$ assuming
dispersion relation (6) for $s=2$ and $1,$ for a \textit{pure
}phase of either 2e- or 2h-CPs, compared to empirical data for 3D
SCs (Nb/Cu and Sn); for two quasi-2D SCs (Y123 and Bi2212 with
$T_{c}\simeq 93$ K and $91$ K, respectively); and a quasi-1D SC
($4$-\AA -wide nanotubes with $T_{c}\simeq
15$ K). The dashed curve labeled $d/s=3/2$ appears in Fig. 6 of Ref. \cite%
{Rosencwaig} and seems to provide the only
adjustable-parameter-free comparison with experimental data in
that paper.\ The ordinate axis refers to a universal ``normalized
order parameter.''\ Data for the 3D and 2D SCs refer to
penetration depth measurements. Nanotube data are gap $\Delta (T)$
measurements giving $\Delta (T)/\Delta (0)$ but are plotted as
$[\Delta
(T)/\Delta (0)]^{2}$ so as to coincide with the 2h-CP condensate fraction $%
m_{0}(T)/m_{0}(0)$\ according to the relation $\Delta
(T)=f\sqrt{m_{0}(T)}$, with $f$ a boson-fermion coupling constant
(which drops out from the normalized order parameter), that
follows for 2h-CP condensates from the generalized BEC theory of
Ref. \cite{Tolma}. The dotted straight line marked $d/s=1/1$
strictly corresponds from (7)\ to $T_{c}\equiv 0$; however, it
serves as a lower bound to all curves with $d/s=(1+\epsilon )/1>1$
for small
but nonzero $\epsilon $, implying \textit{quasi}-1D geometries for which $%
T_{c}>0.$ Also shown for reference are the two-fluid model \cite%
{GorterCasimir}\ curve\ $1-(T/T_{c})^{4}$ and the BCS normalized
gap $\Delta (T)/\Delta (0)$ order-parameter curve \cite{Muhlsch}.

\begin{center}
\textbf{Acknowledgments}
\end{center}
 MdeLl thanks UNAM-DGAPA-PAPIIT
(Mexico) for grant \# IN106401 as well as CONACyT (Mexico) for
grant \# 41302F in partial support. He is also grateful for the
hospitality shown him at the Texas Center for Superconductivity,
University of Houston during a sabbatical leave.

   *E-mail: valenciacvd@yahoo.com.mx

\end{document}